\documentclass[preprint2]{aastex}

\newcommand{\phiprec}{\mbox{$\phi_{35}$}}

\newcommand{\phispin}{\mbox{$\phi_{\mbox{\tiny spin}}$}}
\newcommand{\pspin}{\mbox{$P_{\mbox{\tiny spin}}$}}
\newcommand{\her}{\mbox{Her~X-1}}
\newcommand{\sax}{\mbox{\em BeppoSAX}}
\newcommand{\xte}{\mbox{\em RXTE}}
\newcommand{\hst}{\mbox{\em HST}}

\newcommand{\ginga}{\mbox{\em GINGA}}

\newcommand{\etal}{\mbox{et\ al.\ }}
\newcommand{\dex}[1]{\hbox{$\times\hbox{10}^{#1}$}}

\newcommand{\kev}{\,\mbox{keV}}

\newcommand{\msun}{\,\mbox{$\mbox{M}_{\odot}$}}

\newcommand{\aanda}  {Astr.\ Astro\-phys.\nolinebreak\ }
\newcommand{\aasupp} {Astr.\ Astro\-phys.\nolinebreak\ Suppl.\nolinebreak\ }

\slugcomment{Submitted to the Astrophysical Journal}

\shorttitle{RXTE observations of Hercules X-1}
\shortauthors{M.\,D.\ Still et~al.}

\begin{document}

\title{{\em RXTE} Observations of Hercules X-1 During the July 1998
Short-high State}

\author{Martin Still\altaffilmark{1}}
\affil{NASA/Goddard Space Flight Center, Code 662, Greenbelt, MD~20771 \\
Physics and Astronomy, University of St Andrews, 
North Haugh, St Andrews, \\ Fife KY16~9SS, Scotland}

\author{Kieran O'Brien}
\affil{Physics and Astronomy, University of St Andrews, North Haugh,
St Andrews, \\ Fife KY16~9SS, Scotland}

\author{Keith Horne}
\affil{Physics and Astronomy, University of St Andrews, North Haugh,
St Andrews, \\ Fife KY16~9SS, Scotland \\
Department of Astronomy, University of Texas, Austin TX~78712}

\author{Danny Hudson\altaffilmark{2}, Bram Boroson}
\affil{NASA/Goddard Space Flight Center, Greenbelt, MD~20771}

\author{Saeqa Dil Vrtilek}
\affil{High Energy Astrophysics Division,  Harvard-Smithsonian Center 
for Astrophysics, \\ 60 Garden Street, Cambridge, MA 02138}

\author{Hannah Quaintrell}
\affil{Dept of Physics and Astronomy, The Open University, Milton 
Keynes MK7 6AA, UK}

\author{Hauke Fiedler}
\affil{Institut of Astronomy and Astrophysics, Ludwig-Maximilian 
University, Scheinerstr.~1, D-81679 Munich, Germany}

\altaffiltext{1}{Universities Space Research Association}
\altaffiltext{2}{Joint Center for Astrophysics, University of Maryland, 
Baltimore County, MD~21250}

\begin{abstract}
We present \xte\ monitoring of the eclipsing X-ray binary Hercules X-1
conducted over the short-high state of July 1998.  This was one of the
last  major short-high states before   the source entered an anomalous
low-state of activity.   A comparison with   previous epochs finds  no
evidence  for   special  behavior  during  these    observations.  We
determine orbital  and pulsar spin  periods to facilitate measurements
of $\dot{P}_{\mbox{\tiny   spin}}$  and  $\dot{P}_{\mbox{\tiny  orb}}$
during the subsequent anomalous low state and during the next epoch of
high-state activity.   Spectrally,  the decay of the  short-high state
and concurrent pre-eclipse dips are  consistent with obscuration of  a
central X-ray source  by a cloud of  non-uniform column density.   The
standard model of  a warped  accretion disk  of finite vertical  scale
height fits  the characteristics of  this  absorber well.  Pre-eclipse
dips have durations a factor  of a few  longer than the characteristic
durations  of dips  during main-high  states.  Pulse profile structure
increases in  complexity  towards  the  tail of the   short-high state
suggesting changes in accretion curtain geometry.
\end{abstract}


\section{INTRODUCTION}
\label{sec:intro}

Her~X-1 (Tananbaum \etal 1972) is an eclipsing X-ray binary containing
a pulsar of  1.4  \msun\ and an A7   stellar companion of  2.2  \msun\
(Middleditch \& Nelson 1976; Reynolds \etal 1997). The system displays
behavior on four separate   periodicities -- the pulsar  spin  period
(1.24-s),  the binary  orbit  (1.7-d), a  super  period  of 35-d which
results  from a retrograde-precessing,   warped accretion disk,  and a
beat between the precessional and orbital periods of 1.62-d.  The warp
engine is poorly understood but is likely to be  a combination of both
radiation- and tidally-driven precession  (Papaloizou \& Terquem 1995;
Pringle 1996).

\begin{figure*}
\begin{picture}(0,0)(20,30)
\put(0,0){\includegraphics{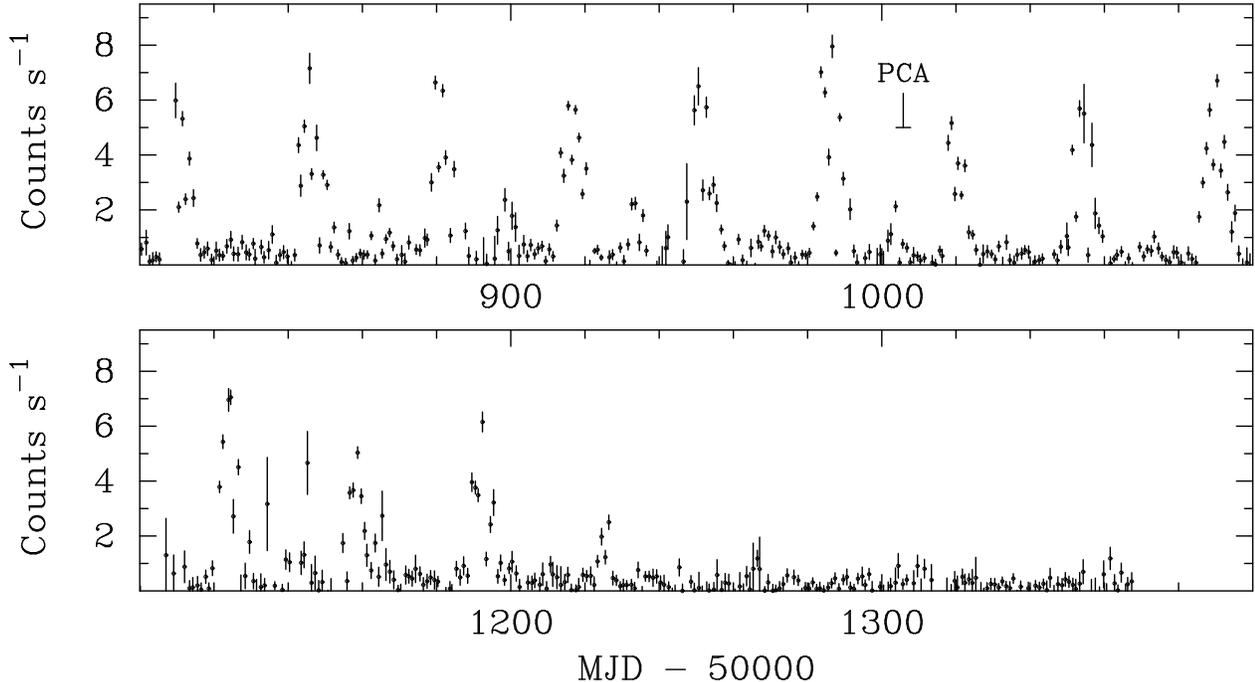}}
\noindent
\end{picture}
\vspace{93mm}
\figcaption[fig1.ps]{The \xte\ ASM light curve of \her. The label ``PCA''
indicates the   time  and combined  duration  of the  current short-high
state visits. \label{asm}}
\end{figure*}

In X-rays the  35-d super-cycle results in two  phases of strong X-ray
activity per  cycle (Giacconi \etal 1973;  Scott \&  Leahy 1999).  The
main-high  has  a  rapid  turn-on  over   $\sim$90-m  and  decays over
$\sim$10-d.  A low-state where  flux  remains at  3--5 percent  of the
high-state level follows   for  the next  $\sim$10-d, succeeded   by a
second   short-high state, lasting   $\sim$5-d,  with flux peaking  at
$\sim$30 percent of the  main-high state maximum.  A further low-state
completes the cycle and extends over the next $\sim$10-d.  UV, optical
and infrared  wavelengths are dominated  by X-ray emission reprocessed
over the inner  face of  the companion star.   There is  no  high--low
state   switching in  these   wavebands because  the  pulsar is always
visible to  a large fraction of the  companion star's surface, however
the accretion disk casts a shadow over  the stellar photosphere, whose
terminator migrates   on the precession timescale.   Consequently, the
super-cycle is  also observed at   UV, optical and infra-red  energies
(Gerend \& Boynton 1976).

The 35-d   clock  has remained  quasi-coherent  since  it's discovery,
although there have  been two recorded   occasions when the  clock has
missed several consecutive turn-ons (Parmar \etal 1985; 1999), and one
occasion when the X-ray flux during main-high states was significantly
reduced (Vrtilek \etal 1994).  The cause of these anomalous low-states
is not  clear, but they probably result  from changes in the  state of
the accretion  disk, either   through  an increase in  vertical  scale
height, or modifications to   the disk warp.  However  the  relatively
constant  level of UV  and  optical flux  during these states suggests
that accretion does not turn off altogether.

During spring  1999, \her\ entered its  latest anomalous low state and
remained there until  the fall of  2000.   Fig.~\ref{asm} presents the
long-term  light curve of \her,  sampled once per  day over the energy
band 2--10\kev,   as obtained by the   \xte\ All-Sky Monitor  (ASM) on
board \xte\ (courtesy  of the ASM team;  Levine et al.   1996).  After
MJD 51200 there are two weak main-high states before  the onset of the
anomalous low-state.  The  data presented in  this paper were taken at
MJD 51000 which  appears to be a normal  short-high state. However all
subsequent short-high states appear either to be absent or weak within
the sensitivity limits of the ASM.  This suggests  that a precursor to
the anomalous   low  state may be  a  weakening   of short-high states
beginning some 200 days  before  the main-high  states  vanish.   This
would have consequences for the timescale of mass transfer variability
in the  system, and the dynamical  reaction  time of  the disk warp to
mass transfer rate onto the neutron star.

The   same  short-high state     was  simultaneously   observed   with
\sax.  Oosterbroek  \etal   (2000) present  data   obtained   with the
Low-Energy  Concentrator  Spectrometer  (LECS;  0.1--10 \kev)  and the
Medium-Energy    Concentrator   Spectrometer  (MECS;    1.8--10  \kev)
experiments over  four orbital  cycles.   They  discern long-duration,
energy-dependent dips in the light  curve, model spectra as composites
of   powerlaw,     blackbody, and    Fe    line   components,  provide
energy-resolved pulse profiles and fit pulse-phased spectral models to
the  data.  The 100~eV blackbody   component displays pulses which are
out of phase with the powerlaw component.  It is believed to originate
from the inner regions of the accretion disk.

Oosterbroek  et  al. (2000) provide  spectral   models and spin-phased
light curves for  the least-absorbed intervals  of each  orbit, but do
not  follow  the spectral  or pulse  profile evolution through dipping
events. \xte's larger collecting area compared with BeppoSAX allows us
to  present  pulse profiles  and spectral   fits through  the  dipping
events, and over an energy band harder than the BeppoSAX LECS and MECS
instruments.    In the   following  paper   we  also  search for   any
outstanding properties of this  short-high state that might signal the
onset   of  any future anomalous low-states,   and  measure the pulsar
orbital and spin  periods to  help  enable the correct  measurement of
orbital and rotational changes during  the low state.  These data form
part of a simultaneous multi-wavelength campaign of \her (Vrtilek
\etal 2001;  Boroson \etal  2000; Boroson \etal  2001;  this paper and
Still \etal  2001), consisting  of pointings   from \xte, the  Extreme
Ultra-Violet Explorer ({\it EUVE}), the Hubble Space Telescope (\hst),
the  4.2-m William Herschel Telescope on  La Palma and the 3.5-m Calar
Alto telescope, Spain.

\section{OBSERVATIONS}

\xte\  pointed  at  \her\   intermittently  during  1998  July  9--13
(MJD~51003.8--51007.9)  accumulating   70-ksec of  time-tagged events.
Assuming   the   35-d  epoch is   determined  by   (M.   Kunz, private
communication):
\begin{equation}
T_{35} = \mbox{MJD}\,50041.0 + 34.85\,E_{35}
\end{equation}
where $T_{35}$ corresponds to X-ray turn-on of the main-high state and
$E_{35}$  is   the cycle number,  the   precession  phases sampled are
$\phiprec$ = 0.63--0.74.   This range corresponds approximately to the
peak   of  a short-high  state to  the   end of  the short-high state,
respectively (Scott \& Leahy 1999).

Event  reconstruction was  performed using standard  algorithms within
{\sc ftools} v4.2. We analyse data from the Proportional Counter Array
(PCA;    Zhang \etal  1993) which  consists     of five identical   Xe
Proportional  Counting Units (PCUs) with  a combined effective area of
6500  cm$^2$.   As well as the  standard  data formats, we  employ two
event  analysers in  the  GoodXenon   event  mode with  2-s   readout.
Timestamps  are  resolved to  1-$\mu$s  while  employing the full  256
energy  channels (1.8--101.0~keV  during epoch  3).  Response matrices
from   1998    July      9  were     obtained  from     the    HEASARC
archive\footnote{http://xte.gsfc.nasa.gov}.  At various  times    3--5
PCUs were active. Background estimates are derived from the Very Large
Event  (VLE)  model to account for   cosmic  events, internal particle
generation             and          South         Atlantic     Anomaly
activation\footnote{http://lheawww.gsfc.nasa.gov/~stark/pca/pcabackest.html}.

\section{LIGHT CURVES}

\begin{figure*}
\begin{picture}(0,0)(20,30)
\put(0,0){\includegraphics{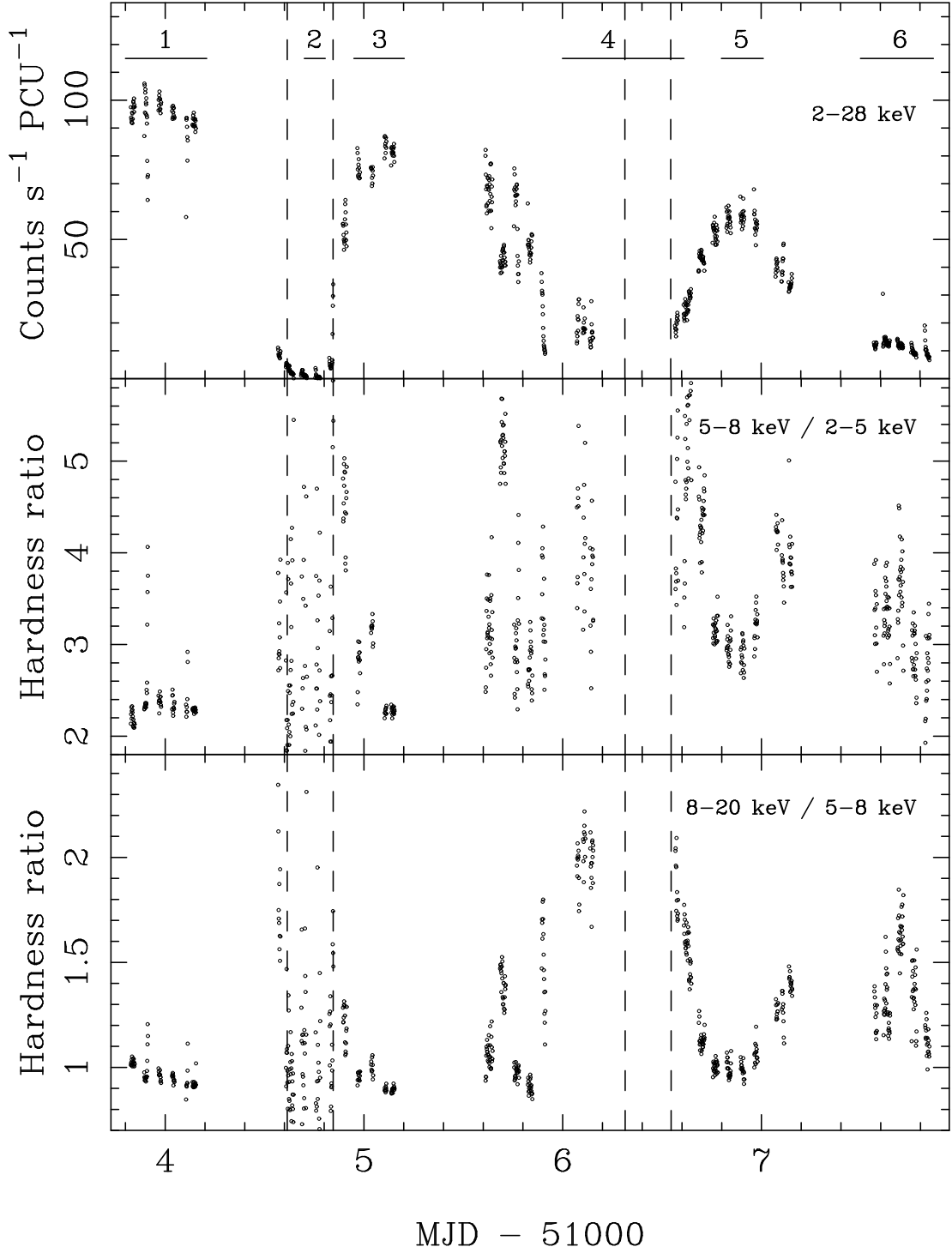}}
\noindent
\end{picture}
\vspace{195mm}
\figcaption[fig2.ps]{The 2--28\kev\ light curve of \her, spanning
approximately the peak of the short-high state to  its end. Dashed lines
correspond to the beginning and end of X-ray eclipse. The lower panels
display the hardness ratios between  the bands 2--5\kev, 5--8\kev\ and
8-20\kev.  Data  used to  extract  spin  pulse profiles (Sec.~6)  were
taken from time sequences 1 to 6. \label{lc}}
\end{figure*}

Standard2  data  from each  visit  were  filtered to reject  pointings
closer than 10 degrees to the Earth's limb and  those off-axis by 0.02
degrees or greater.  Events  were summed across pulse-height  channels
to  provide a light curve  in the  energy  range 2--28\kev\ with 100-s
sampling  and over all PCA  layers and columns.   This is presented in
Fig.~\ref{lc} with   hardness   ratios of 5--8\kev\   /  2--5\kev\ and
8--20\kev\   / 5--8\kev.  Background  models  have been subtracted and
data scaled by the number of active PCUs.

X-ray count rates display a general trend of decline between the first
and last visits, interrupted by  dipping episodes.  The event centered
at MJD 51004.7  is  an eclipse of  the  X-ray source by  the companion
star.  The same eclipse  observed in  the broad  UV emission lines  by
\hst\ provided   constraints on the location   of dynamic  gas  in the
system  (Boroson \etal 2000).  We observe  residual counting events at
mid-eclipse  with   a   weighted-average  rate   of  $1.0   \pm   0.2$
counts\,s$^{-1}$\,PCU$^{-1}$, determined from visits 9  and 10.   From
time-dependent  spectral  fits to  {\em GINGA  LAC} eclipse pointings,
Choi  \etal (1994b) and Leahy  (1995) argue  for an extended, ionized,
scattering region around \her\ producing  these residual photons, from
where the mid-eclipse  fluxes were found to  be variable by a factor 7
from orbit to  orbit. It was suggested that  variability is related to
the    35-d cycle, although  no   correlation  has  been searched for.
However,  by folding spectral models  of  visits 9 and  10 (Section 4)
through the response  matrix of the {\em LAC}  we find these  residual
counts  to be consistent   with  the  {\em  GINGA}  detection  of  2.1
counts\,s$^{-1}$ at $\phiprec = 0.67$.

Between MJD 51005.8--51006.8 we observe an  extended dip with duration
$\sim$ half an orbital period. Although our sampling  is not ideal the
data are suggestive  of dips with  similar phasing and duration during
both the preceding and following orbital cycles.  This is confirmed by
the  more complete sampling of  the contemporaneous \sax\ observations
of  Oosterbroek \etal  (2000).    Dip  ingresses occur earlier    each
consecutive orbit, while the dip  duration increases.  Hardness ratios
suggest   these are, at least   partially, the result of photoelectric
absorption,  although the dips remain   strong features at energies $>
10$\kev.  Both  Shakura \etal (1998) and  Scott \& Leahy (1999) folded
data from the  \xte\ ASM archive and found  that short-high state dips
have significantly longer durations than   main-high state dips.   The
ASM shows the width of these features is coherent  from cycle to cycle
over many years  and therefore is not related  to the  impending X-ray
low-state.  Long dip durations are probably  the result of blending by
numerous unresolved shorter events (Parmar \& Reynolds 1995).

Crosa \& Boynton (1980) determined that dip ingresses migrate linearly
towards earlier orbital  phase over each 35-d  period, where the  time
between successive ingresses is   1.65  days.  Although this  is   not
exactly   the  beat between  the  35-d and  orbital cycles,  Crosa and
Boynton suggest that the dips are the  result of periodic increases in
the scale height of the outer  accretion disk brought  on by bursts of
mass transfer as the $L_1$ point  of the companion star sweeps through
the X-ray shadow of the  disk.  The dip period is  the sum of the beat
period and  the  orbital timescale of structure  in  the outer disk to
travel from the  accretion stream impact  point to the pulsar  line of
sight.   It is difficult  to see how  this model  incorporates dips of
precession-coherent duration  unless the   accretion stream meets  the
disk at a radius that varies over the 35-d cycle. This would result in
a range of orbital frequencies for the newly-arrived gas that broadens
the dip.

Schandl (1996) suggests  the 1.65   day dip  period results from   the
accretion stream  skimming over the surface of  the  disk and crashing
into it at a disk radius and azimuth that vary over the 35-d cycle due
to the disks warped  shape. The  width  of the dips would  most likely
increase as the stream migrated closer to the  disk's center, where an
arbitrary  surface area   subtends a larger   occulting  angle for the
compact   object.  The  detailed  shape  of  Schandl's  model  is  not
consistent with the    inference that the  stream  reaches   the inner
accretion disk  during  the short-high phase of   the cycle,  but this
particular disk  shape also fails  to reproduce the observed X-ray and
optical variability from the source. However, by adopting this type of
model for the  dips,  it may be  possible  to constrain disk  shape by
measuring the duration of dips as a function of 35-d phase.
 
\section{SPECTROSCOPY}

\begin{figure*}
\begin{picture}(0,0)(20,30)
\put(0,0){\includegraphics{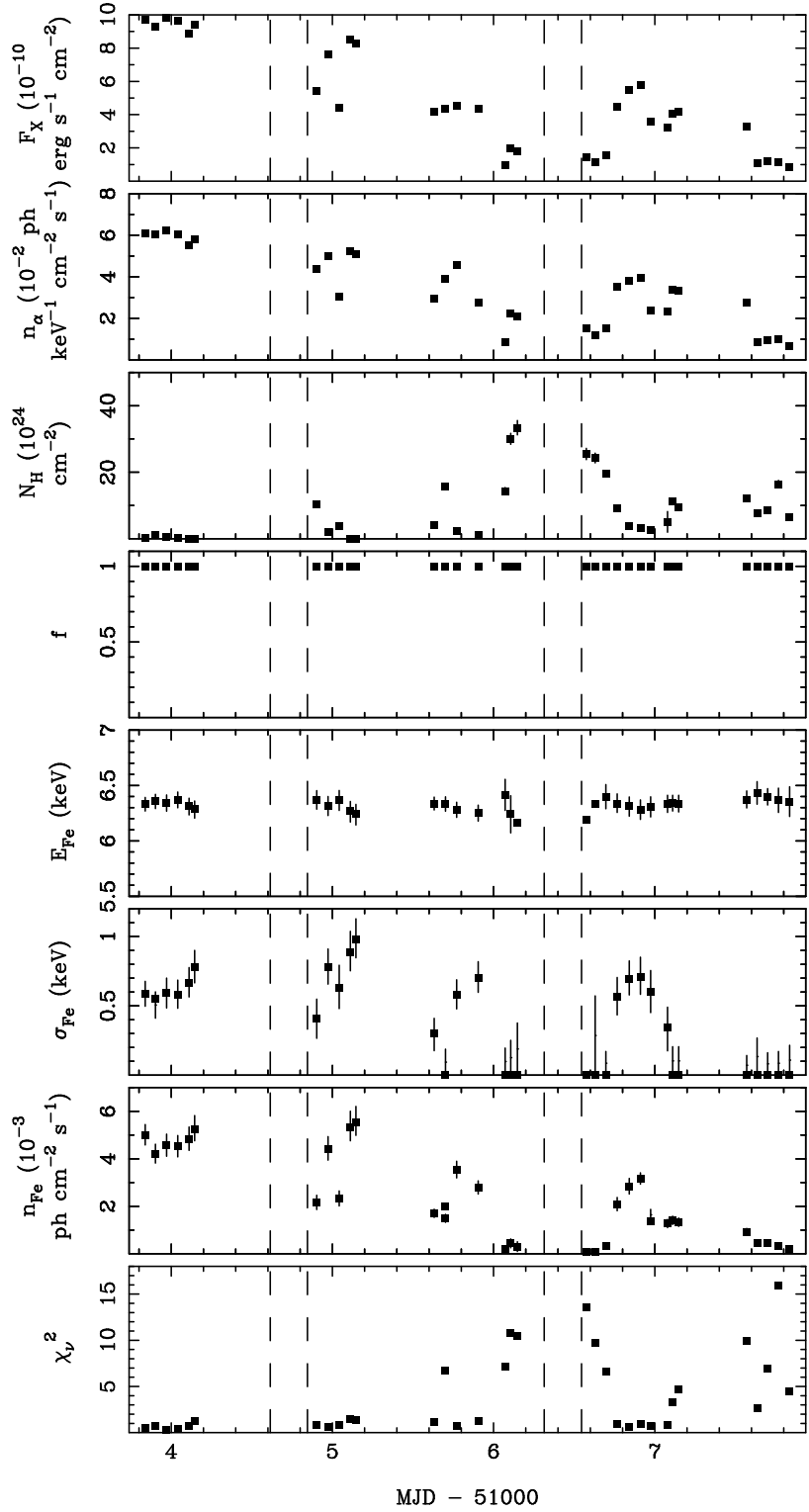}}
\noindent
\end{picture}
\vspace{195mm}
\figcaption[fig3.ps]{Spectral fits for each visit, where 
$F_{\mbox{\tiny X}}$ is the model flux, $\alpha$ a power-law exponent,
$n_\alpha$    the normalization of    the power-law component, $N_H$ a
neutral absorption  column density, $f$  the partial covering fraction
of the absorber, $E_{\mbox{\tiny Fe}}$ the energy of the Fe K feature,
$\sigma_{\mbox{\tiny    Fe}}$   the   width    of  the iron   feature,
$n_{\mbox{\tiny Fe}}$ its normalization  and $\chi^2_\nu$  the reduced
goodness  of   fit measure.   Error bars  are    90 percent confidence
limits. For these fits, $f = 1$ is fixed to represent a total covering
absorber. \label{specfits}}
\end{figure*}

\begin{figure*}
\begin{picture}(0,0)(20,30)
\put(0,0){\includegraphics{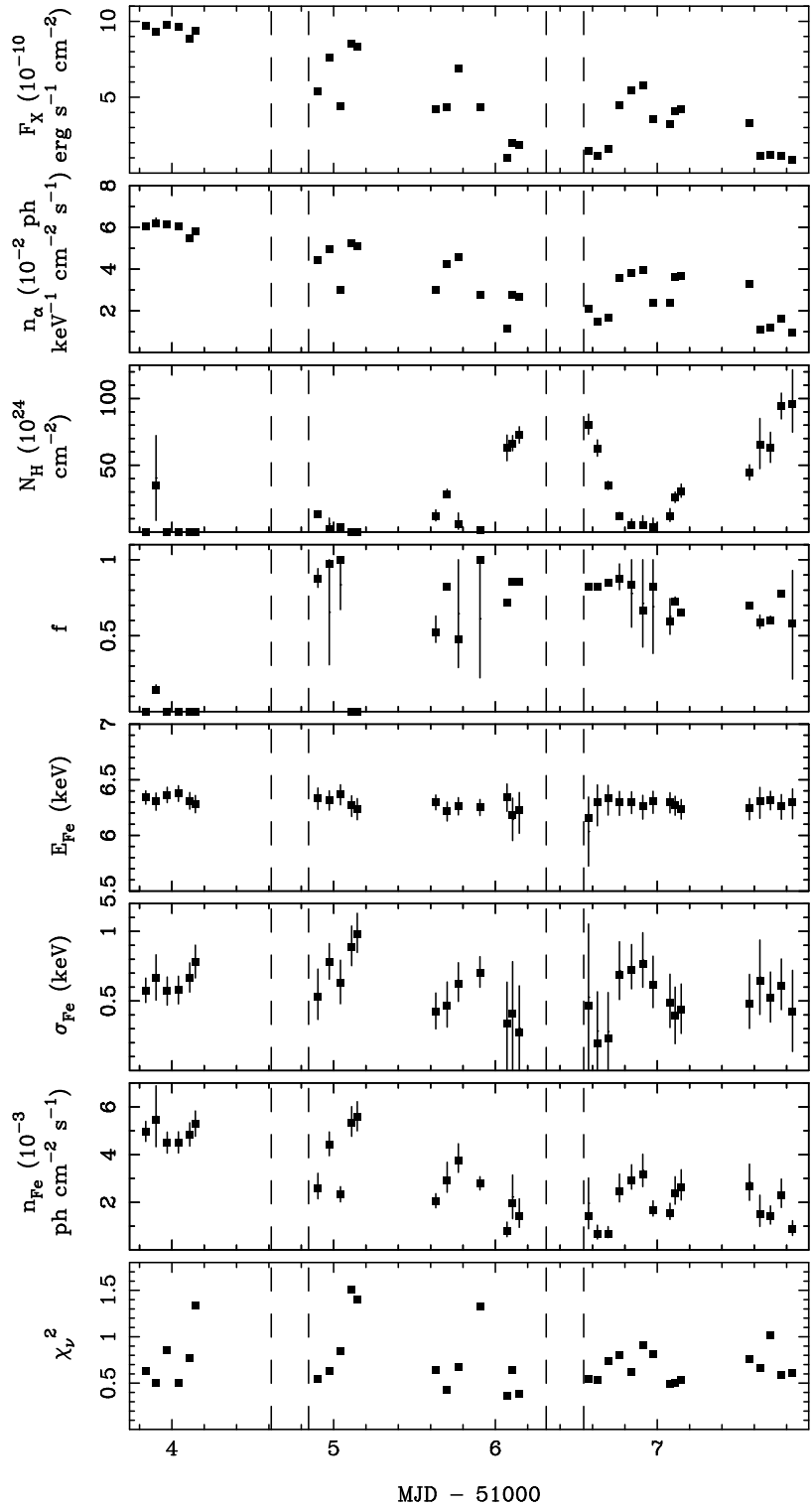}}
\noindent
\end{picture}
\vspace{195mm}
\figcaption[fig4.ps]{As for Fig~\ref{specfits} but for a 
partial-covering absorber, where $f$ is a free parameter. 
\label{specfits2}}
\end{figure*}

\begin{figure*}
\begin{picture}(0,0)(20,30)
\put(0,0){\includegraphics{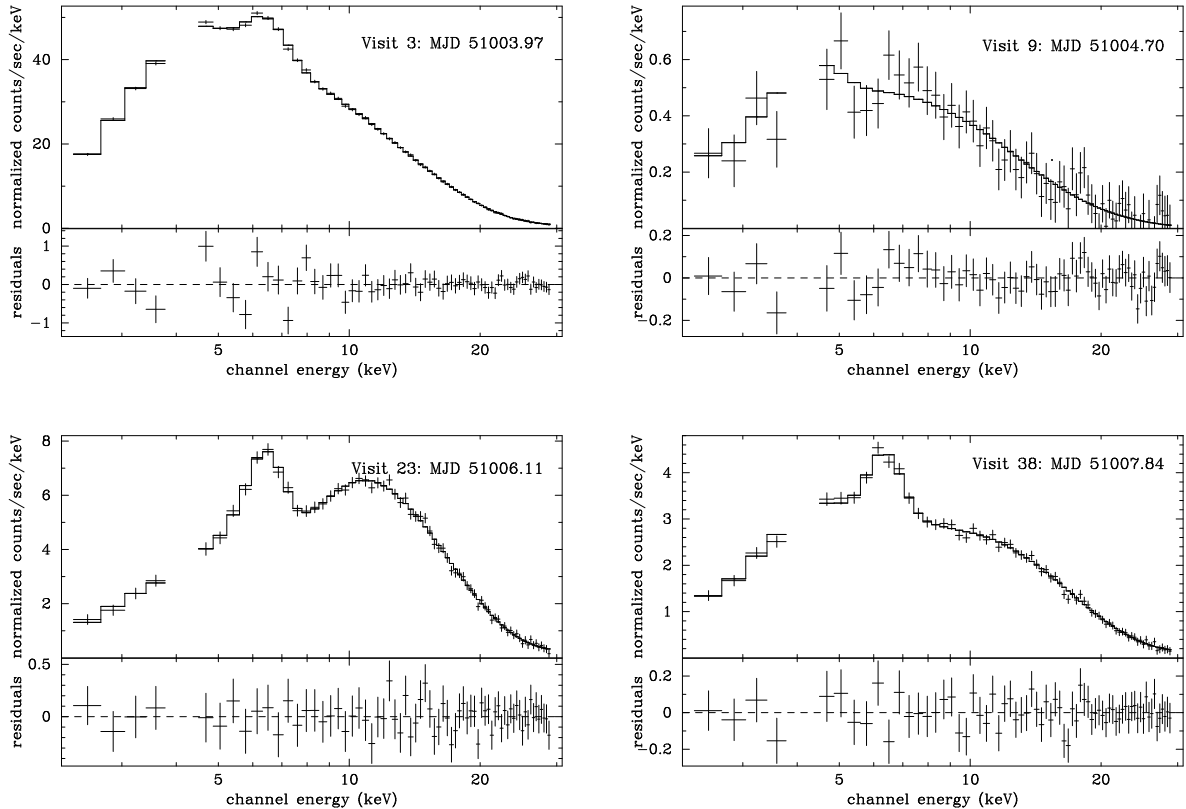}}
\noindent
\end{picture}
\vspace{94mm}
\figcaption[fig5.ps]{Data (crosses) and sample fits (histograms) from 
visits  3  (short-high state  peak intensity),  9  (X-ray eclipse), 23
(broad  dip) and 38 (short-high state  tail).   Residuals for each fit
are in the lower panels. \label{specs}}
\end{figure*}

The current \her\ model predicts that the tail of the short-high state
is the result of the  warped  disk rim  gradually occulting the  X-ray
source as the  disk precesses on the 35-d  cycle (Petterson 1977).  We
expect  both   partial-covering  absorption  from  the  tenuous  upper
atmosphere of the  disk and opaque obscuration from  the thick disk at
smaller heights above the  mid-disk.  Similarly, dips are possibly the
result of periodic  increases in disk height  (Crosa \& Boynton 1980),
absorption  from  the ``bright spot'' impact   site between stream and
disk (Choi  \etal 1994a), or the accretion  stream itself  with ejecta
from the impact splashing above the disk plane (Schandl 1996). In this
section we  test  whether the  picture  above is  consistent with  the
time-varying energy spectrum of \her.

A single absorber is probably unsuitable to fit a system containing an
accretion disk    whose density varies with  vertical height, and with
further possible absorption from winds, coronae and magnetic accretion
columns.  However the data would do a poor  job constraining the large
number of free parameters a more complex  fit requires.  Nevertheless,
we expect the  partial covering model to  provide a better statistical
fit than the homogeneous absorption model.  We  adopt a simple cut-off
powerlaw model with  cold,  blanket absorption and  compare  this to a
similar model with a partial-covering absorber.

Under  the same filtering constraints as  in  section 3, the Standard2
events were sampled with the full  pulse-height resolution of the PCA.
Channels   80--255 (30--100 keV)  were  ignored  as  a result of  poor
counting statistics. Channels 1--6  were ignored due  to uncertainties
in  background modeling.  A   cyclotron absorption  feature  of energy
30--40 \kev\ is usually present in the spectrum of \her\ (e.g.  Mihara
\etal 1990;  Dal  Fiume \etal 1998). The   soft wing of  this feature
encroaches  the PCA  bandpass.   Including  this   wing in the   model
provides acceptable fits  but  since the  center  of the  line is  not
sampled the line parameters afford too much  freedom to the low energy
components of the  model.  We therefore only  fit data softer  than 15
\kev\   to avoid cyclotron  contamination.    Cutting  off the  energy
spectrum below 3 \kev\ avoids  soft contamination from the high energy
tails of spectral components  reprocessed by the inner accretion disk.
These were modeled as a 0.09 \kev\  blackbody component and a broad Fe
L emission complex at 0.95 \kev\ by Dal Fiume \etal (1998).  

\begin{figure*}
\begin{picture}(0,0)(20,30)
\put(0,0){\includegraphics{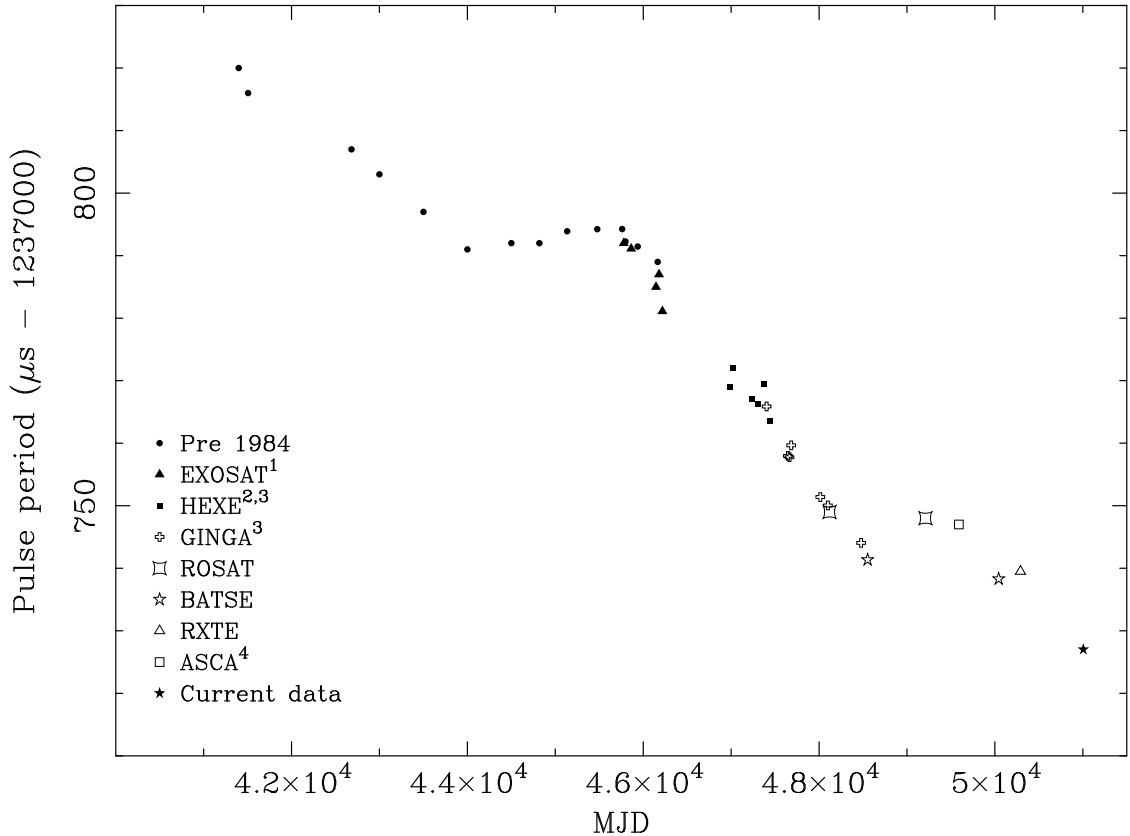}}
\noindent
\end{picture}
\vspace{93mm}
\figcaption[pdot.ps]{The pulse history of \her. (1) Kahabka (1987); (2)
Kunz   \etal  (1996);    (3)  Scott    (1993);  (4)   Stelzer    \etal
(1997). \label{pdot}}
\end{figure*}

\begin{figure*}
\begin{picture}(0,0)(20,30)
\put(0,0){\includegraphics{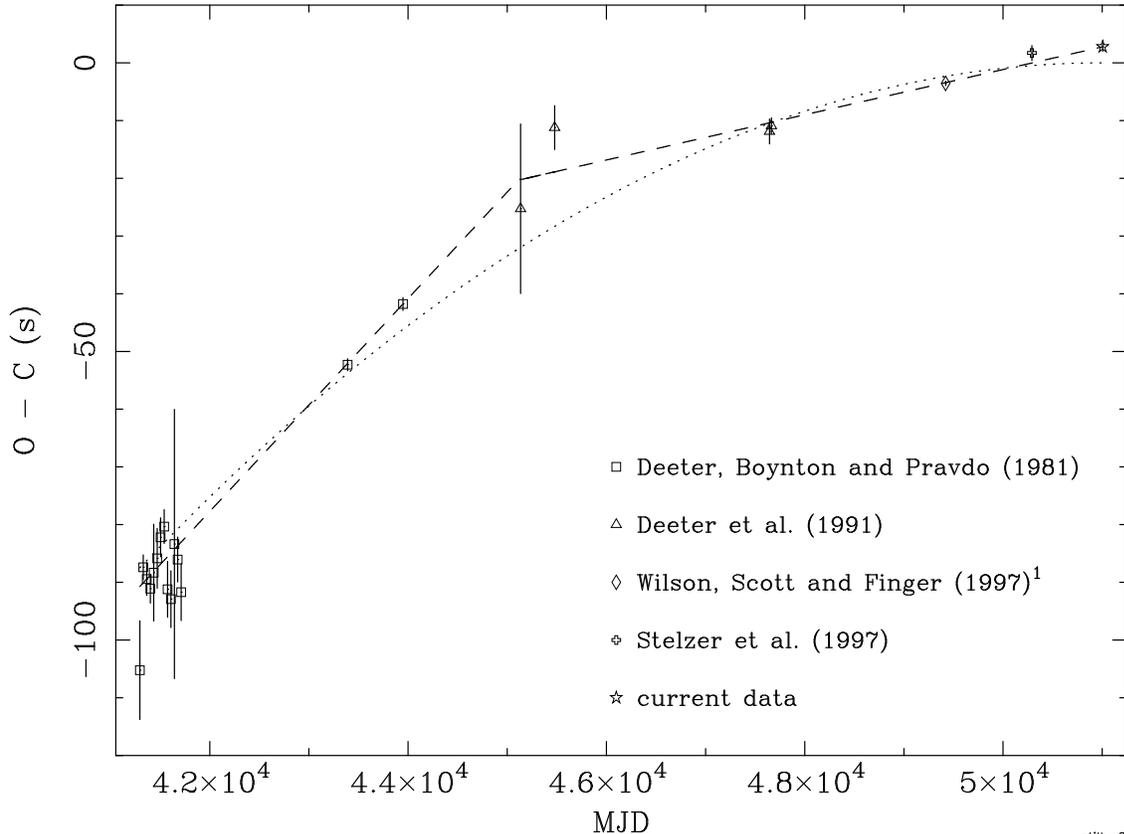}}
\noindent
\end{picture}
\vspace{90mm}
\figcaption[o-c.ps]{Times of superior conjunction of the neutron star 
minus the first  two terms of the  2nd order polynomial fit determined
in   Sec.~5.   The   dotted line   is    the quadratic  term  and  the
discontinuous  line is two linear fits  broken at MJD  45120.  (1) The
mean {\em BATSE} measurement between MJD 48100--50600. \label{o-c}}
\end{figure*}

An interstellar column   density of $N_{H_{\mbox{\tiny int}}}$  =  5.1
\dex{19} cm$^{-2}$, consistent with  the  column to the source  during
the main-high state (Dal Fiume \etal 1998), is  included in the model.
The effective  absorption   cross-section  $\sigma(E)$ was  determined
assuming neutral, solar  abundance material (Baluci\'{n}ska-Church  \&
McCammon 1992).

A broad Fe  K$\alpha$ emission line is detected  at  all times, except
eclipse.  We  model  this with a  Gaussian of   height $n_{\mbox{\tiny
Fe}}$, $\sigma$-width $\sigma_{\mbox{\tiny  Fe}}$  and spectral energy
$E_{\mbox{\tiny Fe}}$.  Fits  to  simple thermal  models of iron  line
plus  absorbed      blackbody or    Bremsstrahlung    emission  proved
statistically unsuitable.

Model fitting employed {\sc  xspec} v10.00.  We consider  the spectral
distribution to be a power law, with  exponent fixed at its unabsorbed
value, $\alpha = 0.9$ with normalization $n_\alpha$, as modeled by Dal
Fiume \etal     (1998)   and   approximately  consistent    with   the
partial-covering   model    used for the   same    short-high state by
Oosterbroek \etal (2000).  This is  subject to  a neutral absorber  of
column density $N_H$ and partial covering  fraction $f$. The composite
model, $S(E)$ is represented algebraically by:
\[
S(E) = e^{-N_{H_{\mbox{\tiny int}}} \sigma(E)}
\left( f e^{-N_H \sigma(E)} + (1 - f) \right) \times
\]
\begin{equation}
\left( n_\alpha E^{-\alpha} 
+ \frac{n_{\mbox{\tiny Fe}} }{\sqrt{2 \pi \sigma_{\mbox{\tiny Fe}}^2}} 
~e^{- \left(\left(  E - E_{\mbox{\tiny Fe}} \right) / 2 \sigma_{\mbox{\tiny Fe}} 
\right)^2 } \right)
\end{equation}

In our first model, where the absorbing column covers the entire X-ray
source, $f = 1$  is fixed as  a  non-variable parameter.  In a  simple
scenario where we  assume the opaque structure  and cool absorber  are
the   lower and  upper parts of    a disk atmosphere   at an arbitrary
distance from the  X-ray source, then  the X-ray flux  $F_{\mbox{\tiny
X}}$ is expected to be correlated with $n_\alpha$, and anti-correlated
with $N_H$ (and $f$ in the partial absorption models).

Fig.~\ref{specfits} summarizes the best-fit  parameters for each visit
after fitting the  blanket-covering model.  The powerlaw normalization
$n_\alpha$ is correlated  strongly with the  overall flux during dips,
eclipses  and ``normal'' states and  would  indicate that much of  the
time-dependent  behavior of \her\ is the  result of opaque obscuration
rather    than   absorption  at   soft  energies.    $N_H$   is indeed
anti-correlated   with $F_{\mbox{\tiny  X}}$.   Goodness  of  fits are
illustrated  in the  lowest  panel  of  the  figure  with the  reduced
$\chi^2$  statistic.  Fits are adequate at  the peak of the short-high
state where there   is  negligible cold absorption but   their quality
deteriorates during dipping events and as the high state decays.  This
verifies that  a blanket-absorber is a  poor model for explaining 35-d
evolution.

Fig.~\ref{specfits2} summarizes the best-fit parameters for each visit
after  fitting the  partial-absorber  model.  We  first  note that the
addition    of    one further  free     parameter,   $f$, has improved
significantly the  $\chi_\nu^2$   from  fits to dip    spectra,  where
$\chi_\nu^2 \sim   1$.  During the peak  of  the  high state, covering
fractions are  approximately 0  with column densities  consistent with
the  interstellar  value.   $N_H$ is  weakly  correlated  with $f$ and
more-strongly  anti-correlated  with  $F_{\mbox{\tiny      X}}$    and
$n_\alpha$.  This indicates  that variability  in $F_{\mbox{\tiny X}}$
is driven by both solid-body obscuration and cold, partial absorption,
at least within the framework  of  this limited  spectral model.   The
reality is plausibly partial-  or fully-covered absorption by a medium
of spatially varying density, as you would  expect from the atmosphere
of an accretion disk.

Within measurement uncertainties the energy of the Fe line is constant
at $6.3  \pm 0.1$ \kev\  over the duration of  the pointings.  This is
slightly lower than  the  measurement of Oosterbroek \etal  (2000) and
\her\ observations  in general.  This  could in  principle result from
erroneously adopting  $\alpha = 0.9$.   However, adopting powerlaws of
various   fixed slope provides no    energy increase and yields either
identical  or  poorer fit  quality.   Additionally,  allowing $\alpha$
freedom   to vary within the  fit   improves $\chi_\nu^2$ but the line
energy does not  increase  significantly.  This  is identical to   the
model   adopted by Oosterbroek \etal  (2000),  and  there is agreement
within uncertainties between  $\alpha = 0.9  \pm 0.1$, $F_{\mbox{\tiny
X}} = 9 \dex{10}$ erg s$^{-1}$  cm$^{-2}$, $\sigma_{\mbox{\tiny Fe}} =
0.6 \pm   0.2$ \kev, $n_{\mbox{\tiny   Fe}} = (2  \pm 1)  \dex{-3}$ ph
s$^{-1}$ cm$^{-2}$ $f  = 1 \pm 0.4$  and  $N_H =  (2 \pm 3)  \dex{19}$
cm$^{-2}$, for $\chi_\nu^2 = 0.9$.

Line strength is correlated with the powerlaw component normalization.
This indicates  that the line strength is  also modulated by both 35-d
effects and dip events.  Dips have the same duration in both continuum
and line.  The inference is that  the powerlaw component and line have
a common locality.

Residual counts remain during  eclipse which can be  fit with a single
powerlaw of  $\alpha  = 1.0 \pm  0.1$  and $n_\alpha = (6.7   \pm 2.0)
\dex{-4}$ ph cm$^{-2}$ s$^{-1}$ keV$^{-1}$ at 1  \kev. The addition of
a line component with a physically suitable  energy between 6--7 \kev\
does not improve the fit statistically.

Fig.~\ref{specs} displays a sample of fits (the maximum and end of the
short-high state, and both mid-eclipse and mid-dip).

\section{PULSE AND ORBITAL EPHEMERIDES}

A measurement  of the   pulse period  at this  epoch  is valuable  for
determining  the spin-up rate  during the low-state using future data.
Fig.~\ref{pdot} displays the neutron star spin period history of \her\
over the previous 30 years.  The general trend  is one of spin-up, but
with at  least  two episodes   of  spin-down at $\sim$~MJD   45000 and
$\sim$~MJD  49000. These  spin-down  episodes  coincide with  the  two
anomalous low  states previously reported by   Parmar \etal (1985) and
Vrtilek \etal (1994).

This indicates  that the anomalous low-states are  related to a change
in the location  of the threading region, where  gas is  stripped from
the  accretion disk by  the  magnetic field  of   the pulsar.  If  the
characteristic distance  of this region is  closer to the neutron star
than the  co-rotation  radius, where  the  rotational velocity  of the
accretion disk exceeds the  rotational velocity of the compact object,
then assuming conservation  of energy and  angular  momentum, the star
will be spun-up. Conversely if the threading region occurs outside the
co-rotation radius,  we expect   spin-down.  Ignoring any  diamagnetic
properties   of the accreting  gas  (King 1993),  the  location of the
threading region is governed by  the balance between magnetic pressure
and the ram pressure  of  the  infall.  Consequently observations   of
spin-down during the anomalous low-states are naturally explained by a
decrease in the  mass transfer rate in  the inner disk. Accretion does
not stop altogether since UV   and optical observations indicate  that
the companion star  is still irradiated strongly   by X-rays from  the
central  object  and inner  disk (Vrtilek  \etal  2001). The continued
brightness at optical and UV  wavelengths also seemingly rules out the
alternative suggestion by  van Kerkwijk \etal  (1998) that spin-torque
reversals  are the result  of  accretion disk warp  flipping the inner
disk over by 180 degrees.

Furthermore Deeter \etal (1991) show that there is also an increase in
the orbital period of the  binary  over time (see Fig.~\ref{o-c})  and
suggest that  a broken linear expression provides  a better fit to the
period measurements than  a continuous one.   The break coincides with
the low-state at $\sim$ MJD 45000,  suggesting that the anomalous lows
also have a measurable effect on the orbital rate. This result is less
convincing than stellar spin-up measurements and requires confirmation
from orbital timing   during, and after, other anomalous   low-states.
Therefore a measurement of the orbital period at this epoch is equally
useful.

For each visit, GoodXenon PCA data were reduced to light curves with a
sample  rate  of 0.02-s  over  the  entire  detector energy  response.
Contemporaneous background models were subtracted and counts scaled by
the number of  active PCUs.  Data with an  elevation above the Earth's
limb less than  10 degrees or offset from the  optical axis by greater
than 0.02 degrees were ignored. Visits 7--11 during X-ray eclipse were
also discarded.  Times were  corrected to the solar system barycenter.
The accuracy of the  on-board clock is 100\,$\mu$s\,d$^{-1}$, although
constant  health   checks  adjust  the   clock  frequency.   Therefore
stability on hourly timescales may be somewhat poorer.

\subsection{Orbital Pulse Delays}

\begin{figure*}
\begin{picture}(0,0)(20,30)
\put(0,0){\includegraphics{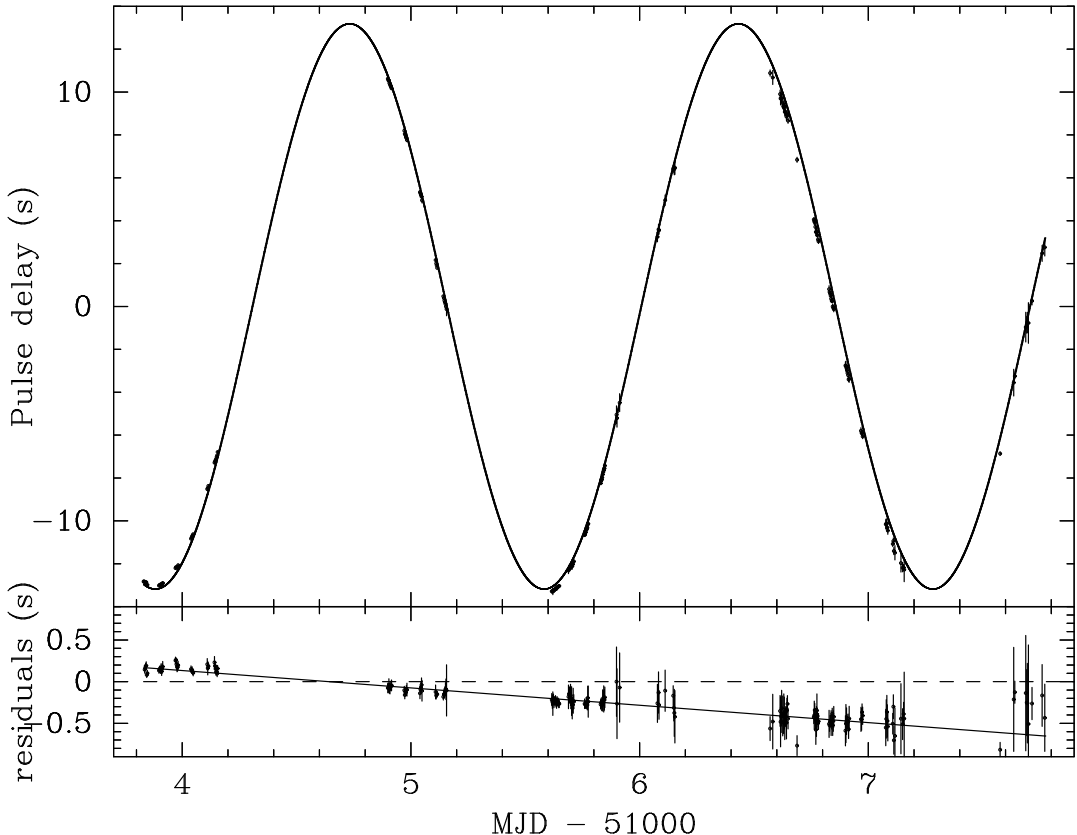}}
\noindent
\end{picture}
\vspace{90mm}
\figcaption[pulsedelay.ps]{Times of pulse maximum arrival relative to 
the center of  mass. The 2nd order  orbital ephemeris of Stelzer \etal
(1997)  is overlayed on the   data.   Residuals plotted in the  lower
panel are accompanied by a linear fit. \label{pulsedelay}}
\end{figure*}

In  order to   measure the  spin   period, we  must  first   determine
time-delays  due to  the  motion  of  the   neutron star.   Data  were
separated  into 200-s intervals  and folded  on  a period close to the
expected pulse    period  during    this epoch,   $P_{\mbox{spin}}   =
1.237730$-s (extrapolated  by eye  from the  BATSE timings  of Wilson,
Scott \& Finger 1997).  A Gaussian function of pulse  phase was fit to
each folded set and the centroid of each fit adopted as pulse maximum.
Due to the 35-d cycle, pulse profiles  vary over the duration of these
visits and consequently this method was more practical than convolving
individual intervals with  each  other.  This provided the  fractional
component of  the orbital   pulse  delay.  The  integer  component was
inferred by  direct comparison with the  orbital fit of  Stelzer \etal
(1997).  The  orbital fit  and  the inferred delays  are  presented in
Fig.~\ref{pulsedelay}.  We    find a linear  trend   over time  in the
residuals  where the  discrepancy between the  true  and assumed pulse
periods results in a gradient  of $-0.219 \pm 0.008$-s\,d$^{-1}$.  The
prediction is  that Fourier  analysis  will reveal a  pulse period  of
$P_{\mbox{spin}} = 1.2377257  \pm 0.0000004$-s after removal  of pulse
delays, where the uncertainty is the 90 percent confidence limit.

The orbital period $P_{\mbox{\tiny orb}}$ was determined by minimizing
$\chi^2$  between   pulse-arrival  times  and a   linear-plus-circular
function  (i.e. Still \etal 1994),  determining the phasing of neutron
star  superior      conjunction,   $T_{\pi/2}$,    and   pulse-arrival
semi-amplitude,  $a  \sin{i}$.  Best  fit parameters  are  provided in
Table~1.

\subsection{Orbital Ephemeris}

\begin{deluxetable}{ccl}
\footnotesize  \tablecaption{System  parameters  determined  from  the
current pulse timings. $T_{\pi/2}$ is defined  as when the mean
longitude equals $\pi/2$ (i.e.  superior   conjunction of the  neutron
star). $a$ is  the separation of the  pulsar and the  binary center of
mass, and $i$ the orbital inclination.
\label{timepar}}
\tablewidth{0pt}
\startdata
\hline\hline
$T_{\pi/2}$ & MJD $51004.729581(9)$ \\
$a \sin{i}$ & $13.1902(9)$-s \\
$P_{\mbox{\tiny orb}}$ & $1.7002(3)$-d \\
$P_{\mbox{\tiny spin}}$ & $1.237727(1)$-s
\enddata
\end{deluxetable}

\begin{deluxetable}{lcccc}
\footnotesize
\tablecaption{Best quadratic and linear fit parameters for the orbital 
ephemeris of \her. The break between the two linear fits occurs at MJD
45120. \label{ephemeris}}
\tablewidth{0pt}
\tablehead{\colhead{} & \colhead{$T_{\pi/2}$ (MJD - 51000)} & 
\colhead{$P_{\mbox{\tiny orb}} (d)$} & 
\colhead{$\dot{P}_{\mbox{\tiny orb}}$ (d\,yr$^{-1}$}) &
\colhead{$\chi^2_\nu$ (d.o.f)}}
\startdata
Quadratic & $4.729549(7)$ & $1.700167427(9)$ & $-1.33(7) \dex{-8}$ & 2.18 (19) \\
Linear 1 & $4.730571(50)$ & $1.700167790(10)$ & -- & 1.13 (14) \\
Linear 2 & $4.729581(7)$ & $1.700167504(7)$ & -- & 1.16 (4)
\enddata
\end{deluxetable}

$T_{\pi/2}$ was combined  with  the previous measurements of   Deeter,
Boynton \& Pravdo (1981), Deeter \etal (1991), Wilson, Scott \& Finger
(1997) and Stelzer \etal (1997).  We fit these  first with a 2nd order
polynomial to determine a new orbital ephemeris,  provided in Table 2,
where
\begin{equation}
T_{\mbox{\tiny ecl}} = T_{\pi/2} + P_{\mbox{\tiny  orb}} E +
\frac{1}{2} P_{\mbox{\tiny  orb}} \dot{P}_{\mbox{\tiny orb}} E^2
\end{equation}
$T_{\mbox{\tiny  ecl}}$  are  the times   of mid-eclipse  and  $E$ the
integer orbital cycle number since $T_{\pi/2}$.

The quality of fit is slightly improved if  we adopt the suggestion by
Deeter  \etal  (1991) that the  distribution is  described by a broken
linear expression.  The  corresponding ephemerides  are also given  in
Table 2 where the break is found to occur at MJD 45120.  Both fits are
overlayed on the  data in the O  $-$ C (Observed $-$ Computed) diagram
of Fig.~\ref{o-c},   where the  first   two  terms  of  the  quadratic
ephemeris have been subtracted from data and fits.

\subsection{Pulse Period}

Orbital motion was  removed  from the  filtered GoodXenon   timings by
subtracting  a  circular function of    semi-amplitude $a \sin{i}$ and
phased by   Linear ephemeris 2 from   Table~2.  A  Lomb-Scargle period
search was performed over the entire sample (Press \etal 1992) and the
spin period determined as $\pspin = 1.237727 \pm 0.000001$, consistent
with  the  prediction from  Sec.~5.1.  The  error is  the $1$-$\sigma$
width  of a   Gaussian  fit to the   power  peak in  the  Lomb-Scargle
statistic distribution.

Fig.~\ref{pdot}  displays the appended  pulse  history where  we see a
general trend of spin-up, with at least two  epochs of spin-down which
coincide with extended X-ray low states (Vrtilek \etal 1994).  We note
that   the first well-sampled episode     of  spin-down at MJD   45000
coincides with the break between  the  two linear orbital  ephemerides
from Sec.~5.2.  \pspin\  time-derivatives  during the three epochs  of
spin-up were determined by linear SVD fits and listed in Table~3.

\begin{deluxetable}{ccc}
\footnotesize
\tablecaption{Linear fits to the three episodes of spin up shown in 
Fig.~\ref{pdot}. \label{pspindot}}
\tablewidth{0pt}
\tablehead{\colhead{Epoch (MJD)} & 
\colhead{$\dot{P}_{\mbox{\tiny spin}}$} ($\mu$s\,y$^{-1}$) &
\colhead{$\chi^2_\nu$ (d.o.f.)}}
\startdata
41399--44000 & -3.81 $\pm$ 0.16 & 1.39 (4) \\
45761--48552 & -6.67 $\pm$ 0.08 & 1.94 (22) \\
49210--51006 & -4.35 $\pm$ 0.27 & 2.41 (3)
\enddata
\end{deluxetable}

\section{PULSE PROFILES}

\begin{figure*}
\begin{picture}(0,0)(20,30)
\put(0,0){\includegraphics{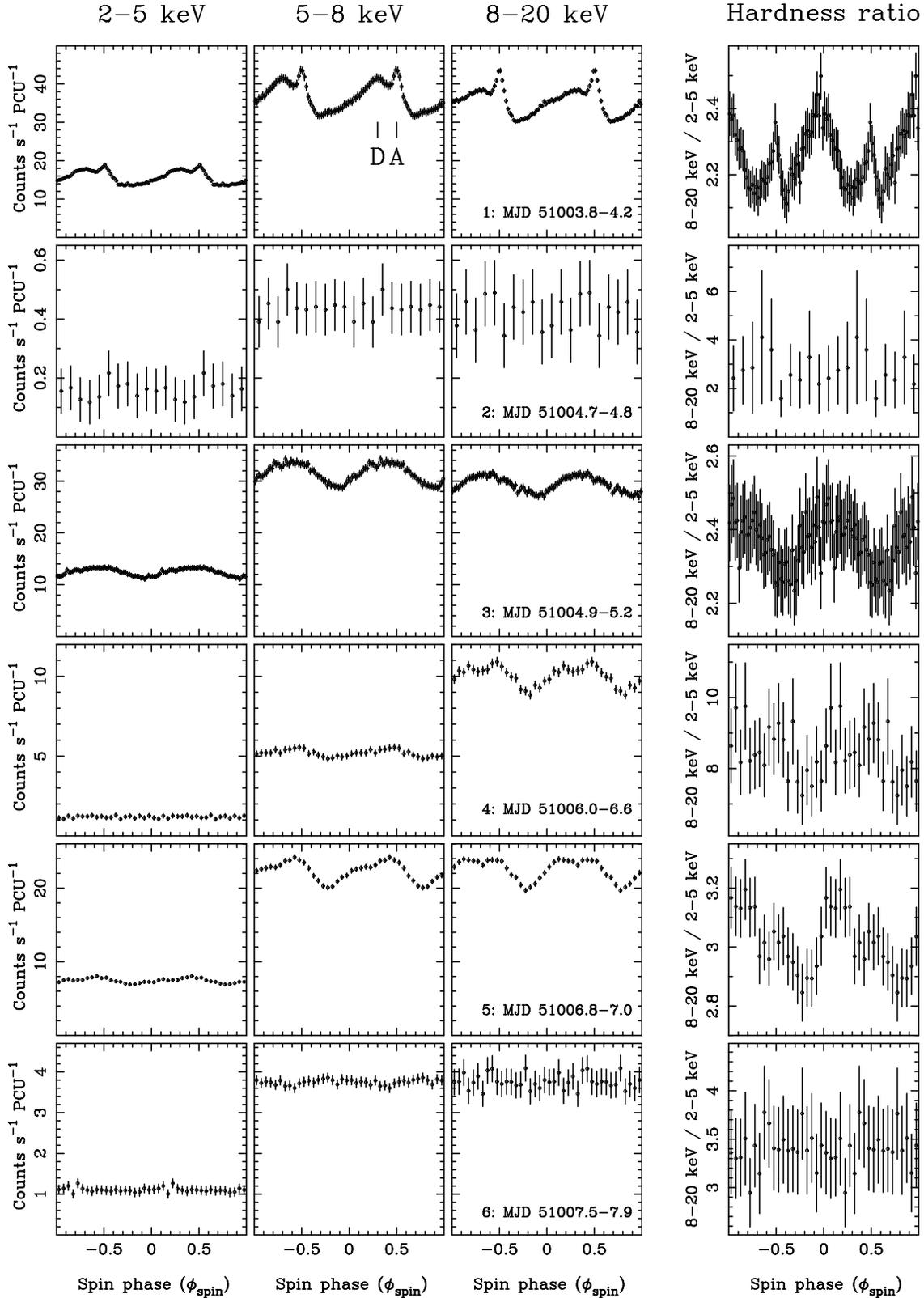}}
\noindent
\end{picture}
\vspace{195mm}
\figcaption[fig9.ps]{Pulse profiles sampled over five time ranges (see
Fig.~\ref{lc}). 1: MJD~51003.8--4.2 (peak of the short-high state), 2:
MJD~51004.7--4.8 (mid-eclipse),  3: MJD~51004.9-5.2 (post-eclipse), 4:
MJD~51006.0--6.6 (dip),   5:  MJD~51006.8--7.0    (post-dip) and    6:
MJD~51007.5--7.9 (the end-points  of the short-high  state), and three
energy bands: 2--5  \kev, 5--8 \kev\ and 8--20  \kev. A and D refer to
pulse    components   identified   by    Scott,   Leahy    \&   Wilson
(2000). \label{pulses}}
\end{figure*}

\begin{figure*}
\begin{picture}(0,0)(20,30)
\put(0,0){\includegraphics{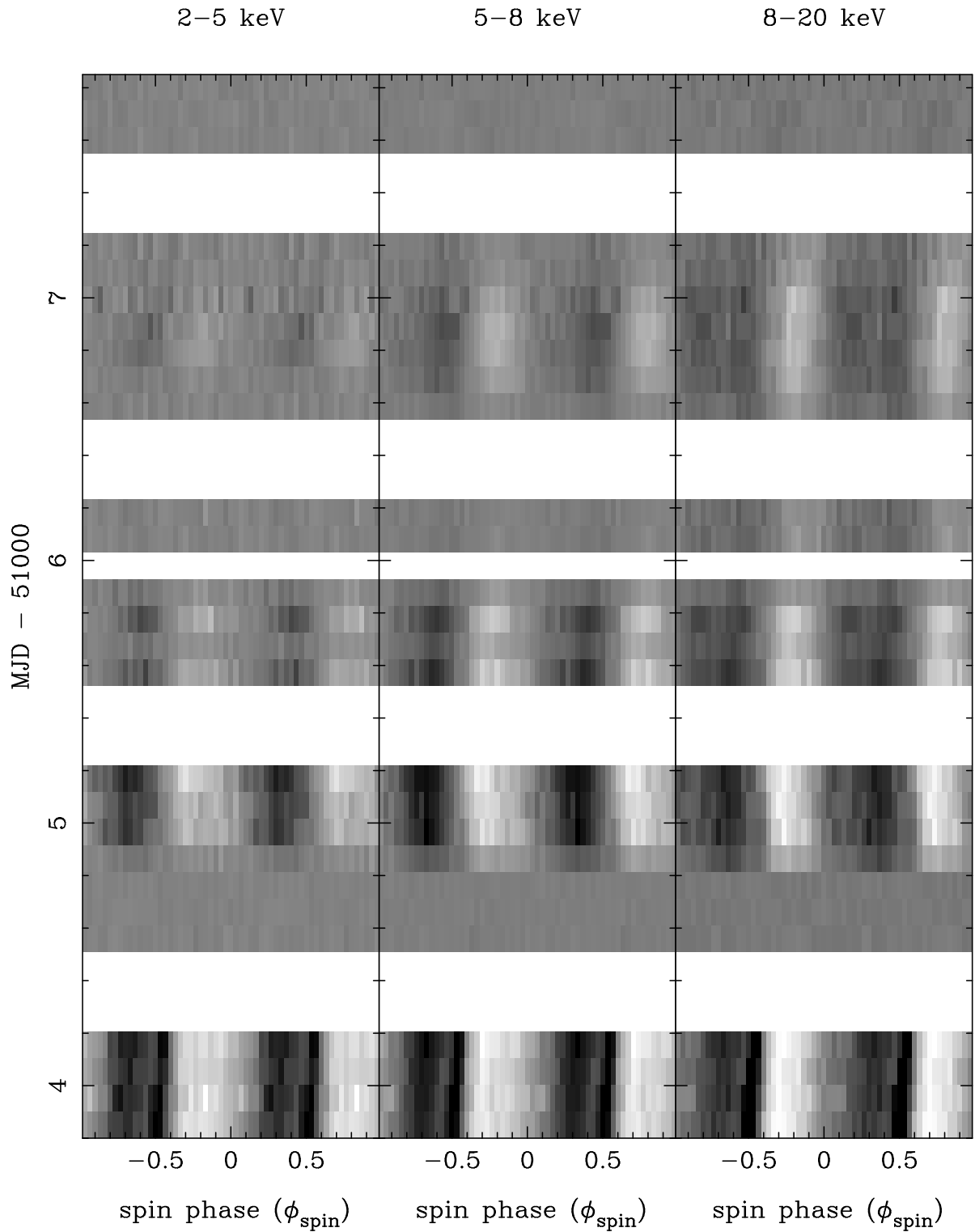}}
\noindent
\end{picture}
\vspace{195mm}
\figcaption[fig10.ps]{2--5 \kev, 5--8 \kev\ and 8--20 \kev\ pulses 
binned uniformly over time, showing the evolution of the pulse profile
from  the  peak of  the  short-high   state to the   beginning of  the
following low-state.  In each individual  bin, the mean count rate has
been subtracted  from the data. White  strips  correspond to time bins
with no data. \label{pulses2}}
\end{figure*}

Finally in this paper we look at the evolution of spin pulses over the
short-high phase. Both Deeter \etal (1998) and Scott, Leahy and Wilson
(2000;  hereafter SLW)  looked in  detail  at the profile and spectral
variability of the short-high state pulses. With this current epoch of
data we are  provided with  short-high  state  sampling improved  over
those previous observations.

We filter the  GoodXenon  events into three  energy bands:  2--5 \kev,
5--8 \kev\ and 8--20 \kev,  remove pulse-delay effects due to  orbital
motion  (Sec.~5.1)  and extract  events  from six separate time ranges
(see Fig.~\ref{lc}): 1:    MJD~51003.8--4.2 (peak of    the short-high
state),  2:   MJD~51004.7--4.8  (mid-eclipse),   3:   MJD~51004.9--5.2
(post-eclipse/pre-dip) 4:  MJD~51006.0--6.6 (dip), 5: MJD~51006.8--7.0
(post-dip)  and 6: MJD~51007.5--7.9 (the  end-points of the short-high
state).  These are presented in Fig.~\ref{pulses}.  Fig.~\ref{pulses2}
presents a smooth   greyscale representation of the pulses  throughout
the decline of the high state.  In this case events have been filtered
into  40 time bins  and 30  phase bins.   Over each time  bin the mean
count rate  has been subtracted from  the data.  Bins which contain no
data are  represented by white  horizontal strips.  Fig.~\ref{pulses2}
shows  a  small phase drift  in  the pulse,   modulated on the orbital
period with amplitude a few  pixels.  This is  most likely caused by a
small error  (within  measurement uncertainties)  in the orbital  time
delay fit.  Consequently  absolute alignment of  features at different
times  is  ambiguous and  caution  is required   when  aligning narrow
features in the pulse profile.

Using  \ginga\ (Deeter   \etal  1998)  and  \xte\   observations,  SLW
developed a geometric model of the X-ray source based on the evolution
of the   pulse profile through the  main-high   and short-high states.
Given an  unobscured  view of the   accretion region,  SLW  argue that
back-scattering off the two magnetic accretion curtains dominates over
forward-scattering.  However, as the  cycle proceeds, the edge of  the
accretion disk passes across our line of sight to the various emission
regions, removing  individual components from  the pulse profile.  SLW
argue that emission from the far curtain (components B and E, in their
terminology) and   emission from   the near  pole   (component C)  are
obscured by the disk at the current 35-d phases (the far pole is never
visible in the SLW model).  We see no obvious signatures of components
B,  C or E  in the observed  profiles, consistent both with this model
and in agreement  with the   \ginga\  satellite observations  of   the
short-high state (Deeter \etal 1998).

The  top  three profiles  of  Fig.~\ref{pulses}, extracted  during the
short-high state maximum, are consistent  with the profiles  presented
by SLW from a similar 35-d phase, showing a narrow peak at \phispin\ =
0.5, superimposed  on a  broad  asymmetric modulation  with minimum at
\phispin\ =  0.7 and  maximum at  \phispin\ =  0.3, thus rising slower
than it falls.  Between the D and A peaks is  a minimum at \phispin\ =
0.6.  It is reasonable to associate the  narrow peak with component A,
which SLW model as emission back-scattered off the near curtain, close
to the magnetic pole on the  surface of the  pulsar.  We associate the
broad peak with component D, which is  emission from the same curtain,
but significantly  higher above the surface  of the star.  Component A
is harder than component D.  Component A  intensity and hardness ratio
are  directly  correlated, while  component   D intensity and hardness
ratio do not appear to be correlated over the whole  pulse cycle.  The
hardness ratio has a  sharp maximum at  \phispin\ = 0.5, corresponding
to the sharp   intensity   peak A,  and  a broad    roughly sinusoidal
modulation peaking at \phispin\ = 0.5,  half way up the intensity rise
of component D.  The dip at  \phispin\ = 0.6 is  not accompanied by an
increase in hardness ratio, as expected from photoelectric absorption.

The  next row  of profiles  are sampled   during  mid-eclipse.  During
mid-eclipse the count rate is 100 times less than  the high state peak
however hardness ratios are essentially unchanged, consistent with the
findings   of Choi \etal   (1994).    Pulse fractions decrease  during
eclipse and there is no compelling  evidence for pulses.  Consequently
the pulses are more deeply eclipsed than the mean light.  The residual
is thought to result from a corona surrounding  the accretion disk and
neutron star and/or the companion star (e.g., Leahy 1995).

The third set of pulse profiles  have been taken from the post-eclipse
recovery. In the time between the first and  third pulses, component A
has  disappeared from the profiles. This  is consistent  both with the
\ginga\ observations in the latter stages  of the short-high state and
the   accretion   disk  occulting the   region   responsible  for this
component.  Component D  is  slightly   harder than before    eclipse,
suggesting a higher column density in front of this region.

Set 4  of  pulse profiles was  extracted from  the center  of the next
absorption dip.  The hardness  ratio has increased  by a factor 3 over
pre-dip pulses, expected from photoelectric absorption, and the dip at
\phispin\ =  0.6 has disappeared.  Here  there are two  peaks at
\phispin\ =  0.2 and \phispin\ =  0.5 and there  is tentative evidence
that both   peaks   are  accompanied by    corresponding  increases in
hardness.  Unlike previous visits,   intensity and hardness  ratio are
directly correlated. No pulses are detected in the 2--5 \kev\ band.

Pulses  extracted   after the dip   and   displayed as  set  5 show  a
predictable  decrease in   hardness  ratio. However   the  correlation
between intensity and  hardness ratio remains, as do  the two  peaks at
\phispin\ = 0.2 and \phispin\ = 0.5.

In the final set of light curves, the mean intensities are a factor 10
smaller than  those in the top  row, with hardness ratios indicating a
larger absorption  column.   Evidence   for pulsations  is    marginal
however, indicating that  the pulse fraction  is reduced by at least a
factor 20.

The model of SLW predicts that the pulse profile should become simpler
as the short-high state evolves into the low state. This is the result
of a number of emission regions being occulted by the disk, one-by-one
as the 35-d cycle proceeds.  This is not  observed.  The complexity of
the pulse profile  increases at the tail-end  of the short-high state.
Since it is not possible for component  A to re-emerge from behind the
accretion disk  at   some time during   MJD~51005.2--5.6, the simplest
interpretation  for  the maximum at \phispin\  =  0.5 is that the pulse
shape of  component  D evolves from   a single  asymmetric  peak  to a
double-peaked form.  Since   the hardness ratios indicate   no obvious
photoelectric  absorption event between  the  two  peaks, perhaps  the
easiest way to achieve this involves a physical change in the geometry
of the  accretion   flow.  This is   a reasonable  solution since  the
location  of the  threading  region between disk  and curtain  will be
modified cyclically with the precession of the accretion disk.

\section{CONCLUSIONS}

We have presented 3--30 \kev\ \xte\ PCA data of what could arguably be
one of  the last normal short-high  states  of the  X-ray pulsar \her\
before it entered an anomalous  low-state of activity.  Fig.~\ref{asm}
illustrates that the \xte\ ASM detected its last major main-high state
before the anomalous  low at   MJD 51195.  There   were  very few  (if
perhaps  no) short-high states  detected  after MJD 51004, although we
note \her\ is a source undergoing deep eclipses and dip events and the
ASM data does not have ideal sampling.  Since accretion disk structure
is projected closer to the X-ray source during short-high states it is
likely that the disappearance of short-high states  may be a precursor
indicating an  impending  anomalous  low-state.  We suggest   that the
event that caused the anomalous low may have began some time before it
was noticed in the ASM.  A study of pulse timings over this period may
reveal that  spin-down began (or  that spin-up was  decelerating) long
before  the high states disappeared.   An  interesting test to perform
would be the     measurement of $\dot{P}_{\mbox{\tiny   spin}}$   more
uniformly over the epoch  covered by Fig.~\ref{asm} using timings from
the {\it BATSE} experiment that was on board {\it CGRO} (e.g., Wilson,
Scott \& Finger 1997).  Comparing the current  epoch PCA data with the
large sample of short-high states  collected by the  \xte\ ASM and the
limited number of previous  pointed  short-high state observations  by
\ginga, we find no evidence for this particular short-high state to be
considered special.

We have determined  the orbital and pulsar spin  periods at this epoch
to  facilitate   measurements   of $\dot{P}_{\mbox{\tiny  spin}}$  and
$\dot{P}_{\mbox{\tiny orb}}$ during the  anomalous low and during  the
next epoch of high-state activity.

Broad-band timing verifies    that  the pre-eclipse  dips  during  the
short-high state   have   durations  longer  than   the characteristic
durations of dips  during main-high states (Scott \&  Leahy 1999).  It
appears likely  that dip durations are  related to the location of the
gas stream impact  with  the  accretion disk,  the trajectory   of the
stream   and the shape   of  the disk  at  the  impact point. Detailed
hydrodynamical calculations may be able  to place model constraints on
these properties using dip observations.

Spectrally, the short-high  state and  the  dips are consistent   with
obscuration  of a central  X-ray source  by a cloud  of varying column
density.  The   standard model of a  warped  accretion disk  of finite
scale height fits this picture well.

Some modification to  the  recent pulse model   of SLW is  required to
explain   the evolution of pulse profiles,   from relatively simple to
more complex, during the tail of the short-high state.  We suggest (as
do SLW) that  geometric changes in the accretion  curtains need to  be
considered to model the pulse profiles correctly.

\acknowledgments 
This  paper   employed \xte\  All  Sky  Monitor results  made publicly
available by the ASM/\xte\ Teams at MIT  and NASA/Goddard Space Flight
Center.  This work was  partially funded by  NASA grants NAG5-6711 and
NAG5-7333. KOB, KH and HQ acknowledge  research grant support from the
UK Particle Physics  and  Astronomy Research Council.  KH acknowledges
support  from  a  Beatrice   Tinsley  Visiting  Professorship   at the
University of  Texas, Austin.  We thank  the   referee for  a thorough
contribution.


\begin{thebibliography}{}

\bibitem[Baluci\'{n}ska-Church and McCammon (1992)]{bal92} 
Baluci\'{n}ska-Church M., McCammon D., 1992, \apj, 400, 699

\bibitem[Boroson \etal (2000)]{bor00} Boroson B., Kallman T., 
Vrtilek S. D., Raymond J., Still M., Bautista M., Quaintrell H., 
2000, \apj, 529, 414

\bibitem[Boroson \etal (2001)]{bor01} Boroson, B. \etal,
2001, \apj, 545, 399

\bibitem[Choi \etal (1994a)]{cho94a} Choi C. S., Nagase F., Makino F., 
Dotani T., Min K. W., 1994, \apj, 422, 799

\bibitem[Choi \etal (1994b)]{cho94b} Choi C. S., Dotani T., Nagase F., 
Makino F., Deeter J. E., Min K. W., 1994, ApJ, 427, 400

\bibitem[Crosa and Boynton (1980)]{cro80} Crosa L., Boynton P. E., 1980,
\apj, 235, 999

\bibitem[Dal Fiume \etal (1998)]{dal98}Dal Fiume D., \etal. 1998, A\&A, 
329, L41

\bibitem[Deeter, Boynton and Pravdo (1981)]{dee81} Deeter J. E., 
Boynton P. E., Pravdo S. H., 1981, \apj, 247, 1001

\bibitem[Deeter \etal\ (1991)]{dee91} Deeter J. E., Boynton P. E., 
Miyamoto S., Kitamoto  S., Nagase F., Kawai  N., 1991, \apj,  383,
324

\bibitem[Deeter \etal\ (1998)]{dee98} Deeter J. E., Scott D. M., 
Boynton P. E., Miyamoto S., Kitamoto S., Takahama, S., Nagase F., 1998,
ApJ, 502, 802

\bibitem[Gerend \& Boynton (1976)]{ger76} Gerend D., Boynton P. E., 
1976, ApJ, 209, 562

\bibitem[Giacconi \etal (1973)]{gia73} Giacconi R., Gursky H., Kellogg 
E., Levinson R., Schreier E., Tananbaum H., 1973, ApJ, 184, 227

\bibitem[Kahabka (1987)]{kah87} Kahabka, P., 1987, PhD-Thesis, Technische 
Univ.\ M\"{u}nchen

\bibitem[King (1993)]{kin93} King A. R., 1993, MNRAS, 261, 144

\bibitem[Kunz et al. (1996)]{kun96} Kunz M. \etal, 1996, \aasupp, 120, 233

\bibitem[Leahy (1995)]{lea95} Leahy D. A., 1995, \apj, 450, 339 

\bibitem[Levine et al.(1996)]{lev96} Levine, A. \etal,
 1996, \apj, 469, L33

\bibitem[Middleditch and Nelson (1976)]{mid76} Middleditch J., Nelson
J. E., 1976, ApJ, 208, 567

\bibitem[O'Brien \etal (2001)]{obr01} O'Brien K., Horne K., Hynes R., 
Chen W., Haswell C., Still M., 2001, MNRAS, submitted

\bibitem[Oosterbroek \etal\ (1999)]{oos99} Oosterbroek T., Parmar A. N., 
Dal Fiume D., Orlandini M., Santangelo A., Del Sordo, S., Segreto A., 2000, 
A\&A, 353, 575 

\bibitem[Nagase (1989)]{nag89} Nagase F., 1989, \pasj, 41, 1

\bibitem[Papaloizou \& Terquem (1995)]{pap95} Papaloizou J. C. B., 
Terquem C., 1995, MNRAS, 274, 987

\bibitem[Parmar \& Reynolds 1995]{par95} Parmar A. N., Reynolds A. P., 1995, 
A\&A, 297, 747

\bibitem[Parmar \etal (1985)]{par85} Parmar A. N., White N. E., Barr
P., Pietsch W., Truemper J., Voges W., McKechnie S., 1985, \nat, 313,
119

\bibitem[Parmar \etal (1999)]{par99} Parmar A. N., Oosterbroek T., 
dal Fiume D., Orlandini M., Santangelo A., Segreto A., del Sordo S.,
1999, A\&A, 350, L5

\bibitem[Petterson (1975)]{pet75} Petterson J. A., 1975, \apj, 201, L64

\bibitem[Petterson (1977)]{pet77} Petterson J. A., 1977, \apj, 218, 783

\bibitem[Press \etal\ (1992)]{pre92} Press W. H., Teukolsky S. A., 
Vetterling W. T., Flannery B. P., 1992, Numerical Recipes, The Art
of Scientific Computing, 2nd edn, CUP: Cambridge

\bibitem[Pringle (1996)]{pri96} Pringle J. E., 1996, MNRAS, 281, 357

\bibitem[Reynolds and Parmar (1995)]{rey95} Reynolds, A. L., Parmar
A. N., 1995, \aanda, 297, 747

\bibitem[Reynolds \etal\(1997)]{rey97} Reynolds A. P., Quaintrell H., 
Still M. D., Roche P., Chakrabarty D., Levine S. E., 1997, MNRAS, 288, 43

\bibitem[Schandl (1996)]{sch96} Schandl S., 1996, \aanda, 307, 95

\bibitem[Scott (1993)]{sco93} Scott M. D., 1993, PhD-Thesis, Univ. 
of Washington

\bibitem[Scott and Leahy (1999)]{Sco99} Scott D. M., Leahy D. A., 1999,
\apj, 510, 974

\bibitem[Scott, Leahy and Wilson (2000)]{sco00} Scott D. M., Leahy 
D. A., Wilson, R. B., 2000, \apj, 539, 392

\bibitem[Shakura \etal\ (1998)]{sha98} Shakura N. I., Ketsaris N. A.,
Prokhorov M. E., Postnov K. A., 1998, MNRAS, 300, 992

\bibitem[Soong \etal (1990)]{soo90} Soong Y., Gruber D. E., Peterson
L. E., Rothschild R. E., 1990, \apj, 348, 641

\bibitem[Stelzer \etal\ (1997)]{ste97} Stelzer B., Staubert R., 
Wilms  J.,   Geckeler R.  D.,  Gruber    D., Rothschild  R., 1997,
Proceedings   of the  Fourth    Compton   Symposium,  C. D.    Dermer,
M. S. Strickman, J. D. Kurfess, AIP, 753

\bibitem[Still \etal (1994)]{sti94} Still M. D., Marsh T. R., Dhillon
V. S., Horne K., 1994, MNRAS, 267, 957

\bibitem[Still \etal (2001)]{still01} Still M. et~al., 2001, 
in press

\bibitem[Tananbaum \etal (1972)]{tan72} Tananbaum H., Gursky H., 
Kellogg E. M., Levinson R., Schreier E., Giacconi R., 1972, ApJ, 174, 
L143

\bibitem[van Kerkwijk \etal (1998)]{van98} van Kerkwijk M. H., 
Chakrabarty D., Pringle J. E., Wijers R. A. M. J., 1998, ApJ, 499,
L27

\bibitem[Vrtilek and Halpern (1985)]{vrt85} Vrtilek S. D., Halpern J. P.,
1985, \apj, 296, 606

\bibitem[Vrtilek \etal\ (1994)]{vrt94} Vrtilek S. D. et~al., 1994, ApJL, 
436, 9

\bibitem[Vrtilek \etal (2001)]{vrt01} Vrtilek S. D., Quaintrell H., 
Boroson B., Still M., Fiedler H., O'Brien K., T., McCray R., 2001, 
ApJ, 548, 471

\bibitem[Wijers and Pringle (1999)]{wij99} Wijers R. A. M. J., Pringle 
J. E., 1999, MNRAS, 308, 207

\bibitem[Wilson, Scott and Finger (1997)]{wil97} Wilson R. B., Scott
D. M., Finger M. H., 1997, Proceedings of  the Fourth Compton Symposium,
C. D.  Dermer M. S. Strickman J. D. Kurfess, AIP, 739

\bibitem[Zhang \etal\ (1993)]{zha93} Zhang W., Giles A. B., Jahoda K., 
Soong Y., Swank J. H., Morgan E. H., 1993, in EUV, X-Ray and Gamma-Ray 
Instrumentation for Astronomy {\sc iv}, O. H. Siegmund, SPIE, Bellingham
WA, 324 

\end{thebibliography}
\end{document}